\documentclass[aps,pre,twocolumn,superscriptaddress,showpacs,amsmath,amssymb,floatfix]{revtex4}

\usepackage{graphicx}
\usepackage{epsfig}
\usepackage{psfig}

\begin{document}

\title{An alternative order parameter for the 4-state Potts model}

\author{H. A. Fernandes\footnote{henrique@pg.ffclrp.usp.br}}
\author{E. Arashiro\footnote{evearash@usp.br}}
\affiliation{Departamento de F\'{\i}sica e Matem\'{a}tica, Faculdade
de Filosofia, Ci\^{e}ncias e Letras de Ribeir\~{a}o Preto,
Universidade de S\~{a}o Paulo, Avenida Bandeirantes, 3900 - CEP
14040-901, Ribeir\~{a}o Preto, S\~{a}o Paulo, Brazil}

\author{A. A. Caparica\footnote{Corresponding author: caparica@if.ufg.br}}
\affiliation{Instituto de F\'{\i}sica, Universidade Federal de
Goi\'{a}s, C.P. 131, 74.001-970, Goi\^{a}nia (GO) Brazil}

\author{J. R. Drugowich de Fel\'{\i}cio\footnote{drugo@usp.br}}
\affiliation{Departamento de F\'{\i}sica e
Matem\'{a}tica, Faculdade de Filosofia, Ci\^{e}ncias e Letras de
Ribeir\~{a}o Preto, Universidade de S\~{a}o Paulo, Avenida
Bandeirantes, 3900 - CEP 14040-901, Ribeir\~{a}o Preto, S\~{a}o Paulo,
Brazil}

\begin{abstract}

  We have investigated the dynamic critical behavior of the
  two-dimensional 4-state Potts model using an alternative order
  parameter first used by Vanderzande [J. Phys. A: Math. Gen.
  \textbf{20}, L549 (1987)] in the study of the Z(5) model. We have
  estimated the global persistence exponent $\theta_g$ by following
  the time evolution of the probability $P(t)$ that the considered
  order parameter does not change its sign up to time $t$. We have
  also obtained the critical exponents $\theta$, $z$, $\nu$, and
  $\beta$ using this alternative definition of the order parameter and
  our results are in complete agreement with available values found in
  literature.

\end{abstract}

\pacs{64.60.Ht, 75.10.Hk, 02.70.Uu}

\maketitle

\section{INTRODUCTION}

In 1989, Janssen \textit{et al.} \cite{Janssen1989} and Huse
\cite{Huse1989} pointed out, using renormalization group techniques and
numerical calculations, respectively, that there is universality and
scaling behavior even at an early stage of the time evolution of
dynamic systems without conserved order parameter (model A in the
terminology of Halperin \textit{et al.} \cite{Halperin1974}).

Since then, a great deal of works on phase transitions and critical
phenomena using Monte Carlo simulations in the short-time regime have
been published and their results are in good agreement with
theoretical predictions and numerical results found in equilibrium
\cite{Zheng1998,Arashiro2003,Okano1997,Dasilva2002,Fernandes2005,
  Santos2001,Grandi2004,Pleimling2004,Pleimling2005,Malakis2003}. In
addition, the new approach has proven to be useful in determining with
good precision the dynamic exponent $z$, as well as the new exponent
$\theta$ which governs the so-called critical initial slip
\cite{Li1995}, the anomalous behavior of the magnetization when the
system is quenched to the critical temperature $T_c$.

The dynamic scaling relation obtained by Janssen \textit{et al.}
\cite{Janssen1989} for the \textit{k}th moment of the magnetization,
extended to systems of finite size, is written as
\begin{equation}
M^{(k)}(t,\tau,L,m_0)=b^{k\beta/\nu}M^{(k)}(b^{-z}t,b^{1/\nu}\tau,b^{-1}L,b^{x_0}m_0),
\label{Eq:scaling}
\end{equation}
where $t$ is the time evolution, $b$ is an arbitrary spatial scaling
factor, $\tau=(T-T_c)/T_c$ is the reduced temperature and $L$ is the
linear size of the lattice. The exponents $\beta$ and $\nu$ are as
usual the equilibrium critical exponents associated respectively with
the order parameter and the correlation length and $x_0$ is related to the
exponents $z$, $\theta$, $\beta$ and $\nu$ by the equation
\begin{equation}
x_0=\theta z + \beta/ \nu.
\end{equation}

By setting the scaling factor $b=t^{1/z}$ and $\tau=0$ in Eq.
(\ref{Eq:scaling}), we obtain the first moment of the magnetization
\begin{equation}
M(t)\thicksim m_0t^\theta, \label{Eq:theta}
\end{equation}
where $m_0$ represents the initial magnetization of the system.

Far from equilibrium, another dynamic critical
exponent was proposed by Majumdar \textit{et al.} \cite{Majumdar1996}
studying the behavior of the global persistence probability $P(t)$
that the order parameter has not changed its sign up to the time $t$.
At criticality $P(t)$ is expected to decay algebraically as
\begin{equation}
P(t) \thicksim t^{-\theta_g},
\end{equation}
where $\theta_g$ is the global persistence exponent. If the time
evolution would be a Markovian process, then the exponent
$\theta_g$ should obey the equation \cite{Majumdar1996}
\begin{equation}
\theta_g z =-\theta z+\frac{d}{z}-\frac{\beta}{\nu}.
\end{equation}
However, as shown in several works \cite{Majumdar1996,Majumdar2003,
  Schulke1997,Dasilva2003,Dasilva2005,Ren2003,Albano2001,Saharay2003,
  Hinrichsen1998,Sen2004,Zheng2002} the exponent $\theta_g$ is an
independent critical index closely related to the non-markovian character
of the process.

In this paper, we estimate the global persistence exponent of the
4-state Potts model employing an alternative order parameter first
used by Vanderzande \cite {Vanderzande1987} in the study of the Z(5)
model. Our estimate is in complete agreement with the results obtained
recently for the Ising model with three-spin interactions in one
direction and for the 4-state Potts model \cite{Condmat2005} with its
traditional order parameter (see Eq. (\ref{Eq:old})). In addition,
comparing with estimates for the Ising and 3-state Potts models
\cite{Schulke1997}, there are consistently increasing values from $q=2$ to $q=4$.
In order to check the validity of that alternative order parameter, we
also estimate the dynamic critical exponents $\theta$ and $z$, along with
the static critical exponents $\nu$ and $\beta$. Our results
[$\theta=-0.046(9)$, $z=2.294(3)$, $\nu=0.669(6)$, and
$\beta=0.0830(6)$] are in complete agreement with previous results
found in the literature. In Section \ref{section2} we describe the
model and the order parameter.  In Section \ref{section3} we show the
short-time scaling relations and present our results. In Section
\ref{section4} we summarize and conclude.

\section{THE MODEL} \label{section2}

The \textit{q}-state Potts model \cite{Potts1952,Wu1992} is a
generalization of the Ising model that preserves the next-nearest
neighbour interaction, works with only two energies (neighboring spins are in
the same state or not) but permits to put at each
site any number of states ($0 \leq q < \infty$). This model encloses a
quite number of other problems of statistical physics. It undergoes a
first-order phase transition when $q > 4$ and a continuous phase
transiton for $q \leq 4$. Thus along with the Ising model, the Potts
model is an important laboratory to check new theories and algorithms
in the study of critical phenomena. Although its exact solution is not
known, several results were obtained during the last fifty years
\cite{Wu1992,denNijs1979,Baxter1982,Domany1978}.

The Hamiltonian of the \textit{q}-state Potts model is given by
\begin{equation}
\beta\mathcal{H}=-K\sum_{\langle i,j\rangle}\delta_{\sigma_i\sigma_j},
\end{equation}
where $\beta=1/k_BT$ and $k_B$ is the Boltzmann constant, $\langle i,j
\rangle$ represents nearest-neighbor pairs of lattice sites, $K$ is
the dimensionless ferromagnetic coupling constant and $\sigma_i$ is
the spin variable which takes the values $\sigma_i=0,\cdots,q-1$ on
the lattice site $i$. It is well known that the critical point of this
model is given by \cite{Wu1992}
\begin{equation}
K_c=\mbox{log}(1+\sqrt{q}).
\end{equation}
Usually, the order parameter of this model is given by
\cite{Zheng1998,Dasilva2004}
\begin{equation}
  M_1(t)=\frac{1}{L^d(q-1)}\left\langle \sum_i(q\delta_{\sigma_i(t),1}-1)\right\rangle
\label{Eq:old}
\end{equation}
where $L$ is the linear size of the lattice and $d$ is the dimension
of the system.

In this paper, we use a different definition for the order parameter,
first proposed by Vanderzande \cite{Vanderzande1987} studying the Z(5)
model. It can be written as
\begin{equation}
  M_2(t)=\frac{1}{L^d}\left\langle \sum_i(\delta_{\sigma_{i}(t),0}-\delta_{\sigma_{i}(t),1})\right\rangle,
  \label{OP}
\end{equation}
where the average $\langle \cdot\cdot\cdot \rangle$ is taken over independent
initial configurations.

\section{RESULTS} \label{section3}

We performed short-time Monte Carlo simulations to obtain the critical
exponents for the 4-state Potts model.

Simulations were carried out for square lattices with periodic
boundary conditions and dimensions $L=120$, 180 and 240. We also used
the lattice sizes $L=20$, 30, 40, 50, 60 and 90 just to estimate the
exponent $z$ through the scaling collapses for different lattice
sizes. The estimates for each exponent were obtained from five
independent bins in the critical temperature. For the exponents
$\theta_g$, $z$, $\beta$ and $\nu$, each bin consisted of 20000
samples, whereas for the exponent $\theta$ we have used 100000
samples. When estimating the exponent $z$ through the scaling
collapses, we used 50000 samples. The error bars are fluctuations of
the averages obtained from those bins. The dynamic evolution of the
spins is local and updated by the heat-bath algorithm.

In the following sections we show the results for the dynamic and
static exponents of the 4-state Potts model.

\subsection{The dynamic critical exponent $\theta_g$}

First of all we are concerned with the global persistence
probability $P(t)$. It is defined as the probability that the global
order parameter has not change its sign up to time $t$. For $\tau=0$,
the global persistence probability decays algebraically as \cite{Majumdar1996}
\begin{equation}
P(t)\thicksim t^{-\theta_g},
\end{equation}
where $\theta_g$ is the global persistence exponent.

In order to estimate the critical exponent $\theta_g$, a sharp
preparation of the initial states is demanded in order to obtain a
precise value for the initial magnetization \hspace{1ex} $m_0 \ll 1$. After
obtaining the critical exponent $\theta_g$ for several values of the
initial magnetization $m_0$, the final value is achieved from
the limit $m_0 \rightarrow 0$.

In this moment, it is worth to explain how obtain a small value of
$m_0$ in Eq. (\ref{OP}). First, each site on the lattice is occupied
by a spin variable which takes the values $\sigma=0$, 1, 2 or 3 with
equal probability.  After, the magnetization is measured by using
$M_2(t)$ and then, the variables in the sites are randomly chosen up
to obtain a null value for the magnetization. The last procedure is to
change $\delta$ sites on the lattice in order to obtain the desired initial
magnetization. It is given simply by
\begin{equation}
m_0=\frac{\delta}{L^2} \label{m_02}
\end{equation}
and a value of $m_0$ is obtained changing $\delta$ sites occupied by
$\sigma=2$ or 3 and substituting them by $\sigma=0$.

\begin{figure}[!ht]
\centering \epsfig{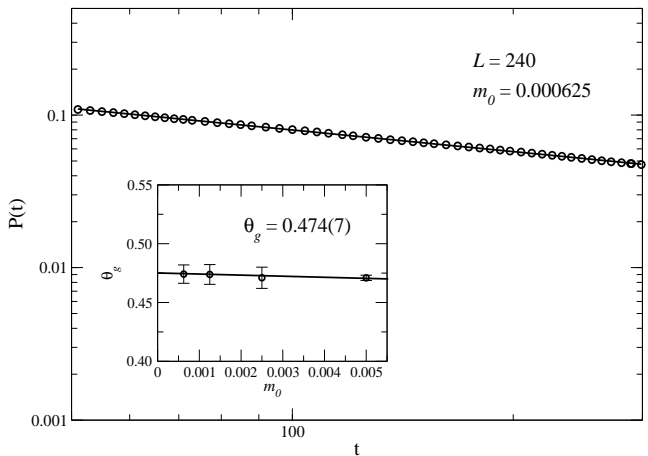}
\caption{The time evolution of the global persistence probability
  $P(t)$ for a lattice size $L=240$ and $m_0=0.000625$. The error bars
  calculated over 5 sets of 20000 samples are smaller than the
  symbols. The inset displays the exponent $\theta_g$ for four
  different initial magnetizations, as well as its extrapolated
  values.}
\label{Fig:persistence}
\end{figure}

In Fig. \ref{Fig:persistence} we show the behavior of the global
persistence probability for $L=240$ and $m_0=0.000625$ in double-log
scales, together with the behavior of the exponent $\theta_g$ for
$m_0=$ 0.005, 0.0025, 0.00125 and 0.000625. In order to obtain these
initial magnetizations for this lattice, $\delta$ should corresponds
to $\delta=288$, 144, 72, and 36.

The extrapolated value of $\theta_g$, when $m_0 \rightarrow 0$ and
$L=240$ is
\begin{equation}
\theta_g=0.474(7).
\end{equation}

The extrapolated values of $\theta_g$ for $L=120$ and $L=180$ are
shown in Table \ref{table:persistence}. We emphasize that sometimes
the value of the initial magnetizations for the lattice size $L=180$
are slighty different from those of the $L=240$. This is necessary in
order to obtain an integer value of $\delta$.

\begin{table}[!ht]\centering
\caption{Global persistence exponent $\theta_g$.}\label{table:persistence}
\begin{tabular}{c c}
  \hline\hline ~~~$L$~~~~ & $\theta_g$~~~\\ \hline
  ~~~120~~~~ & 0.470(5)~~~\\
  ~~~180~~~~ & 0.473(6)~~~\\
  ~~~240~~~~ & 0.474(7)~~~\\ \hline\hline
\end{tabular}
\end{table}

The estimates obtained with the three lattice sizes show that the
finite size effects are less than the statistical errors. Therefore,
we conclude that the values for an infinite lattice are
within the error bars of our results for $L=240$.

\subsection{The dynamic critical exponent $\theta$}

Another critical exponent found only in the nonequilibrium state is
the exponent $\theta$ that characterizes the anomalous behavior of the
order parameter in the short-time regime. Formerly, a positive value
was always associated to this exponent \cite{Zheng1998,Jaster1999,
  Schulke1995,Tome1996,Tome1998,Alves2003,Fernandes2005} and the
phenomenon was known as critical initial slip. However, some models
can exhibit negative values for the exponent $\theta$. This is the
case, for instance, of the Baxter-Wu model \cite{Arashiro2003} and
the tricritical Ising model, anticipated by Janssen \textit{et al.}
\cite{Janssen1994} and numerically confirmed by da Silva \textit{et
  al.} \cite{Dasilva2002}.

In this paper we reobtain the dynamic critical exponent
$\theta$ for the 4-state Potts model using the order parameter
described in Section \ref{section2} (Eq. (\ref{OP})).

Usually the exponent $\theta$ has been calculated
using Eq. (\ref{Eq:theta}) or through the autocorrelation
\begin{equation}
A(t) \thicksim t^{\theta-\frac{d}{z}}, \label{Eq:autocorrelation}
\end{equation}
where $d$ is the dimension of the system. In the present
work however we estimated the exponent $\theta$ using the time
correlation of the magnetization \cite{Tome1998}
\begin{equation}
C(t)=\langle M(0)M(t)\rangle,
\label{Eq:correlation}
\end{equation}
which behaves as $t^\theta$ when $\langle M(t=0) \rangle=0$. The
average is taken over a set of random initial configurations. Initially,
this approach had shown to be valid only for models which exhibit
up-down symmetry \cite{Tome1998}. Nevertheless, it has been
demonstrated recently that this approach is more general and can
include models with other symmetries \cite{Tome2003}. This approach
can thus be used for the $q \neq 2$ Potts models.

When compared to the other two techniques (Eq. (\ref{Eq:theta}) and
Eq.(\ref{Eq:autocorrelation})), this method has at least two
advantages. It does not demand a careful preparation of the initial
configurations ($m_0 \ll 1$) neither a delicate limit $m_0 \rightarrow
0$, as well as the knowledge in advance of the exponent $z$ (see Eq.
(\ref{Eq:autocorrelation})), which is an order of magnitude greater
than $\theta$. In this case, a small relative error in $z$ induces a
large error in $\theta$.

\begin{figure}[ht]
\centering \epsfig{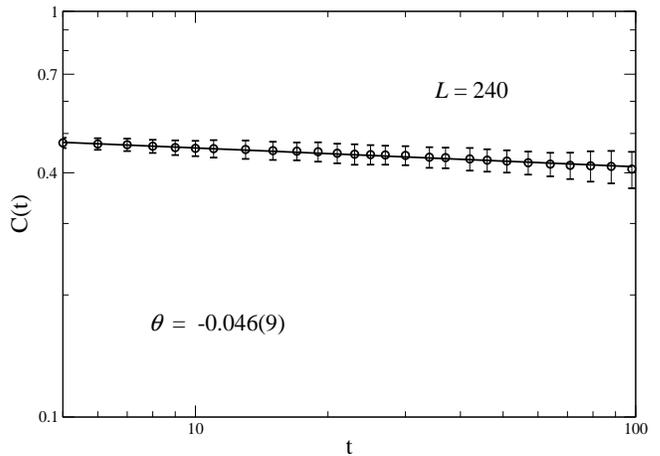}
\caption{The time correlation of the total magnetization for samples
with
  $\langle M(t=0)\rangle =0$. The error bars were calculated over 5
  sets of 100000 samples.}
\label{Fig:correlation}
\end{figure}

In Fig. \ref{Fig:correlation} we show the time dependence of the time
correlation $C(t)$ in double-log scales for the system with $L=240$.
The linear fit of this curve leads to the value $\theta=-0.046(9)$.

For the lattice sizes $L=120$ and $L=180$ we obtained respectively
$\theta=-0.045(8)$ and $\theta=-0.046(8)$. These results are in good
agreement with those found for the same model using the order
parameter of the Eq. (\ref{Eq:old}) \cite{Dasilva2004,Condmat2005}.

\subsection{The dynamic critical exponent $z$}

The critical exponent $z$ was estimated independently by means of two
techniques. We began using mixed initial conditions, in order to
obtain the function $F_2(t)$, given by \cite{Dasilva2002b}
\begin{equation}
F_{2}(t)=\frac{\langle M^2(t) \rangle_{m_0=0}}{\langle M(t)
\rangle^2_{m_0=1}}\thicksim t^{d/{z}}, \label{Eq:f2}
\end{equation}
where $d$ is the dimension of the system. This approach proved to be
very efficient in estimating the exponent $z$ for a great number of
models \cite{Dasilva2002,Arashiro2003,Alves2003,
  Dasilva2004a,Fernandes2005,Dasilva2005}. In this technique, for
different lattice sizes, the double-log curves of $F_2$
\textit{versus} $t$ fall on the same straight line, without any
rescaling of time, resulting in more precise estimates for $z$.

\begin{figure}[ht]
\centering \epsfig{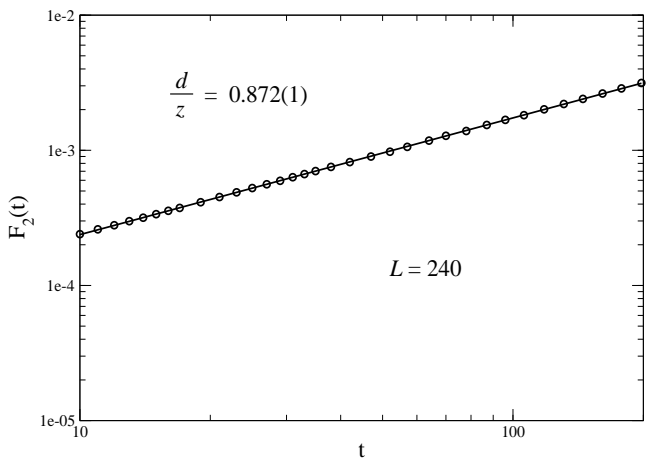}
\caption{The time evolution of $F_2(t)$. The error bars are smaller
than the symbols. Each point represents an average over 5 sets of
20000 samples.} \label{Fig:f2}
\end{figure}

The time evolution of $F_2$ is shown on log-scales in
Fig. \ref{Fig:f2} for $L=240$.

The slope of the straight line gives
\begin{equation}
\frac{d}{z}=0.872(1),
\end{equation}
yielding
\begin{equation}
z=2.294(3).
\end{equation}
This \hspace{1ex} result \hspace{1ex} is \hspace{1ex} in \hspace{1ex}
good agreement with the values
$z=2.294(6)$ recently obtained for the Baxter-Wu model \cite{Arashiro2003},
$z=2.3(1)$ for the Ising model with multispin interactions \cite{Simoes2001}, and
$z=2.290(3)$ for the 4-state Potts model \cite{Dasilva2002b}.

For $L=120$ we have obtained $z=2.296(5)$ and for $L=180$ we have
obtained $z=2.295(5)$ indicating that the finite size effects
are less than the statistical errors.

The second technique consists of studying the parameter
\begin{eqnarray}
  R(T,t,L)=\biggl{\langle} \left( \mbox{sign}\frac{1}{L}\sum_{top}(\delta_{\sigma_{i}(t),0}-\delta_{\sigma_{i}(t),1})\right) \times \nonumber \\ \left( \mbox{sign}\frac{1}{L}\sum_{bottom}(\delta_{\sigma_{i}(t),0}-\delta_{\sigma_{i}(t),1})\right) \biggr{\rangle}
\end{eqnarray}
introduced by de Oliveira \cite{Deoliveira1992}. In this case, the
scaling relation for $T=T_c$ is given by \cite{Soares1997}
\begin{equation}
R(T=T_c,t,L_1)=R(T=T_c,b^{-z}t,bL_1), \label{Eq:R}
\end{equation}
with $b=L_2/L_1$. This equation shows that the dynamical exponent $z$
can be easily estimated by adjusting the time rescaling factor
$b^{-z}$ in order to obtain the best scaling collapse of the curves
for two different lattice sizes.

\begin{figure}[ht]
\centering \epsfig{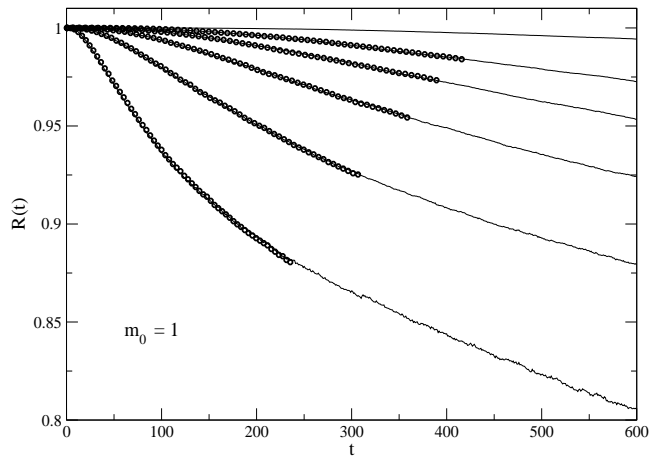}
\caption{$R(t)$ vs $t$ with ordered initial configurations $m_0=1$.
  The full lines show the behavior of $R(t)$ for lattices $L=20$, 30,
  40, 50, 60 and 90 (from the bottom to the top) and the corresponding
  time rescaled curves for lattices $L=30$, 40, 50, 60, and 90 (open
  circles). The exponent $z$ obtained for each collapse is shown in
  Table \ref{Table:collapse}. The error bars, calculated over five
  sets of 50000 samples, are smaller than the symbols.}
\label{Fig:r1}
\end{figure}

Fig. \ref{Fig:r1} shows the parameter $R$ as a function of the time
(full lines), as well as the scaling collapse (open circles) between
different pairs of lattice for samples with ordered initial
configurations $(m_0=1)$.

The best values of $z$, obtained through the $\chi^2$ test
\cite{Numerical} for different scaling collapses are shown in Table
\ref{Table:collapse}.
\begin{table}[!ht]\centering
  \caption{Estimates of the dynamical exponent $z$ for the best scaling collapse of $R(t)$.}\label{Table:collapse}
\begin{tabular}{c c}
\hline\hline ~~~$L_2 \longmapsto L_1$ &~~~~ $z$~~~ \\
\hline
~~~$30 \longmapsto 20$ &~~~~ 2.27(5)~~~ \\
~~~$40 \longmapsto 30$ &~~~~ 2.28(4)~~~ \\
~~~$50 \longmapsto 40$ &~~~~ 2.28(5)~~~ \\
~~~$60 \longmapsto 50$ &~~~~ 2.29(3)~~~ \\
~~~$90 \longmapsto 60$ &~~~~ 2.28(3)~~~ \\
\hline\hline
\end{tabular}
\end{table}

Our results obtained for the collapse of $R(t)$ are in good agreement
with our results arising from $F_2(t)$, as well as the results for the
4-state Potts model \cite{Dasilva2002b}, and the Baxter-Wu model which belongs
to the same universality class \cite{Arashiro2003}.

\subsection{The static critical exponents $\nu$ and $\beta$}

\begin{figure}[ht]
\centering \epsfig{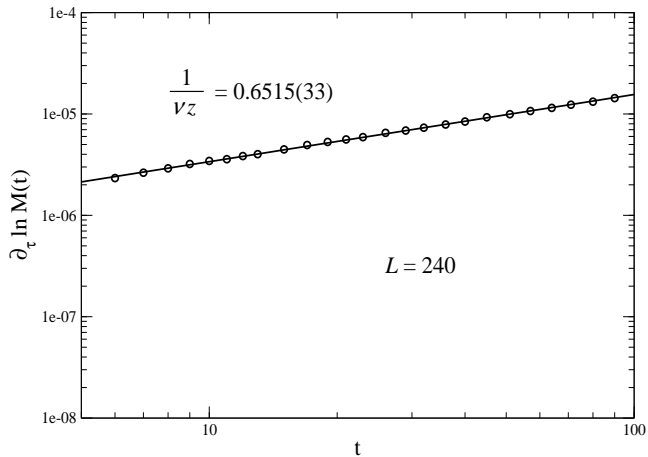}
\caption{The time evolution of the derivative $\partial_\tau
\mbox{ln}M(t,\tau)|_{\tau=0}$ on log-log scales in a dynamic process
starting from an ordered state $(m_0=1)$. The error bars are smaller
than the symbols. Each point represents an average over 5 sets of
20000 samples.} \label{Fig:derivative}
\end{figure}

With the value of the exponent $z$ in hand, we can estimate the static
exponent $\nu$ taking the derivative of the logarithm of the order parameter
\begin{equation}
M(t,\tau)=t^{-\beta/\nu z}M(1,t^{1/\nu z}\tau) \label{Eq:magnetization}
\end{equation}
with respect to $\tau$ in the critical point
\begin{equation}
\partial_\tau \mbox{ln}M(t,\tau)|_{\tau=0}=t^{1/ \nu z}
\partial_{\tau'} \mbox{ln}F(\tau')|_{\tau'=0}.
\label{Eq:derivative}
\end{equation}
This equation follows a power law that does not depend on $L$ and the
function $F(\tau')$ is a scaling function which modifies the power law
at $\tau' \neq 0$ but still in the critical domain.  In numerical
simulations we approximate the derivative by a finite difference. Our
results were obtained using finite differences of $K_c\pm\delta$ with
$\delta=0.001$. In Fig.  \ref{Fig:derivative} the power law increase
of Eq.  (\ref{Eq:derivative}) is plotted in double-log scales for
$L=240$.

From the slope of the curve we estimate the exponent $1/\nu z$ for the
three lattice sizes. Using the exponent $z$ calculated previously, we
obtain $\nu=0.670(9)$ for $L=120$, $\nu=0.668(6)$ for $L=180$, and
$\nu=0.669(6)$ for $L=240$.

\begin{figure}[!ht]
\centering \epsfig{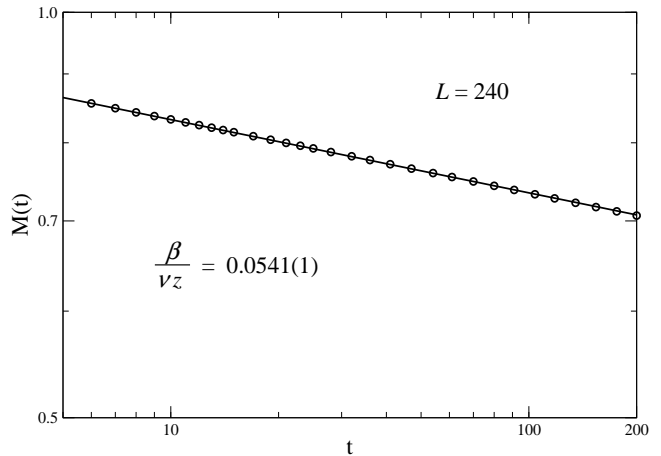}
\caption{The time evolution of the magnetization for initially
ordered samples $(m_0=1)$. The errorbars calculated over 5 sets of
20000 samples, are smaller than the symbols.}
\label{Fig:magnetization}
\end{figure}

Finally, we evaluate the static exponent $\beta$ following
the decay of the order parameter in initially ordered samples
($m_0=1$). At the critical temperature $\tau=0$, the scaling law of
Eq. (\ref{Eq:magnetization}) allows one to obtain $\beta/\nu z$ which
in turn leads to the exponent $\beta$, using the previous result
obtained for the product $\nu z$. In Fig. \ref{Fig:magnetization} we
show the time evolution of the magnetization in double-log scale for $L=240$.

A linear fit of this straight line gives the value $\beta/\nu
z=0.0541(1)$ leading to $\beta=0.0830(6)$. For $L=120$
we obtained $\beta=0.0834(7)$ and for $L=180$ we obtained $\beta=0.0835(4)$.

Our results for $\nu$ and $\beta$ are in good agreement with the exact
results $\nu=2/3$ and $\beta=1/12$ \cite{Baxter1982,Domany1978}.

\subsection{Anomalous dimension $x_0$}

Finally, we calculate the value of the anomalous dimension $x_0$ of
the magnetization which is introduced to describe the dependence of
the scaling behavior of the initial conditions. It is related to the
exponents $\theta$, $z$, and $\beta / \nu$ by
\begin{equation}
x_0=\theta z + \beta / \nu.
\end{equation}

Table \ref{Tb:x0} shows the values of $x_0$ obtained with the
exponents estimated all along this paper.
\begin{table}[!htb]\centering
  \caption{The exponent $x_0$ for the 4-state Potts model.}
  \label{Tb:x0}
\begin{tabular}{c c}
  \hline\hline
  ~~~ L & ~~~~~ $x_0$ ~~~ \\
  \hline
  ~~~ 120 & ~~~~~ $0.021(21)$ ~~~ \\
  ~~~ 180 & ~~~~~ $0.019(20)$ ~~~ \\
  ~~~ 240 & ~~~~~ $0.019(23)$ ~~~ \\
  \hline\hline
\end{tabular}
\end{table}

Our results show that the anomalous dimension of the 4-state Potts
model has a null value whose meaning is the presence of the marginal
operator, i.e., the operator which has the scaling dimension equal to
dimensionality of the system and whose effect is not modified under
renormalization-group operations. An unlike value was found recently by
Arashiro \textit{et al.} \cite{Condmat2005} for this model and for
the Ising model with three-spin interactions in one direction.

\section{DISCUSSION AND CONCLUSIONS} \label{section4}

In this paper we revisited the 4-state Potts model in order to obtain
the global persistence exponent $\theta_g$ using an order parameter
first proposed by Vanderzande \cite{Vanderzande1987} in the study of
the Z(5) model. The results are in good agreement with each other. By
using this alternative order parameter, we have also estimated the
dynamic critical exponents $\theta$ and $z$, as well as the well-known
statical exponents $\nu$ and $\beta$. The exponent $\theta $ was
estimated using the time correlation of the magnetization, whereas to
obtain the exponent $z$ we used the function $F_2(t)$ which combines
simulations performed with different initial conditions and scaling
collapse for the parameter $R$ introduced by de Oliveira. The statical
exponents were obtained through the scaling relations for the
magnetization and its derivative with respect to the temperature at
$T_c$. Our results, when compared with available values in literature
support the reliability of this new order parameter.

\section*{ACKNOWLEDGMENTS}

 This work was supported by the Brazilian agencies CAPES, FAPESP and
FUNAPE-UFG.

\end{document}